\documentclass[12pt]{book}

\usepackage[dvips]{graphicx,color}
\usepackage{makeidx,tsukuba}

\makeauthorindex
\makeindex

\begin{document}

\BookTitle{\itshape The 28th International Cosmic Ray Conference}
\CopyRight{\copyright 2003 by Universal Academy Press, Inc.}
\pagenumbering{arabic}

\chapter{
Searching for a Long Cosmic String Through the Gravitational 
Lensing Effect.}


\author{%
%
%
Yuji SHIRASAKI,$^1$
Ei-ichi MATSUZAKI,$^2$
Yoshihiko MIZUMOTO,$^1$
Fumio KAKIMOTO,$^2$
Syoichi OGIO,$^2$
Naoki YASUDA,$^1$
Masahiro TANAKA,$^1$
Hideki YAHAGI,$^1$
Masahiro NAGASHIMA,$^1$ and
George KOSUGI$^3$ \\
{\it
(1) National Astronomical Observatory of Japan, Mitaka, Tokyo, 181-8588, Japan\\
(2) Department of Physics, Tokyo Institute of Technology, Meguro, Tokyo, Japan\\
(3) Subaru Telescope, NAOJ,  Hilo, HI 96720, USA}
}

\section*{Abstract}

It has been suggested that cosmic strings produced at a phase transition
in the early universe can be the origin of the extremely high energy
cosmic rays (EHCR) observed by AGASA above 10$^{20}$ eV.
Superheavy cosmic strings with linear mass density of 10$^{22}$ g/cm can
be indirectly observed through the gravitational lensing effect the
distant galaxies. The lensing effect by a long straight object can be
characterized by a line of double galaxies or quasars with angular
separation of about 5 arcsec.
We have searched for aligned double objects from the archived data
taken by the Subaru Prime Focus Camera (Suprime-Cam). The Suprime-Cam has a
great advantage in observing the wide field 
of view (30$\times$30 arcmin$^{2}$) with high sensitivity (R$<$26 400s
exposure), so it is suitable for this research.
In this paper, we describe the result of simulation study for developing
the method of searching the objects lensed by cosmic strings, and
present the observational result obtained by this method.

\section{Introduction}

The cosmic ray with energies higher than 10$^{19}$ eV is believed to
be originated out of our galaxy, since no concentration is observed
toward the galactic plane~[5].
In addition to this, the cosmic rays which exceed the GZK cut
off are observed by AGASA experiment~[4], which
suggests that the source is at distance closer than 100 Mpc.
No prominent active sources, however, are found toward the direction
where the extremely high energy cosmic (EHCR) rays are detected, so
the decay of heavy relics from early universe, such as a cosmic string,
has been considered as one of the candidates of origin of the
EHCR~[1].
The standard theory of particle physics predicts a symmetry breaking
in the early universe and, as a result, the production of topological 
defects~[7].
If the defects were generated at the GUT energy, the cosmic string
can be observed as the origin of gravitation lensing~[6]
and its detection 
can be an observational confirmation of the standard theory.
The recent observations of cosmic microwave background
radiation rule out pure topological defects model as the
origin of large scale structure of the universe~[3], however, they still do
not rule out the existence of the defects.
So it is of great importance to constrain the existence experimentally.

In this paper, we propose a method to search for gravitational lensing
by a cosmic string, especially by a long straight string.
The method is applied to the actual data and the results are presented.

\section{Method}

\begin{figure}
   \parbox[b]{0.5\textwidth}{
      \centerline{
         \includegraphics[width=0.45\textwidth]{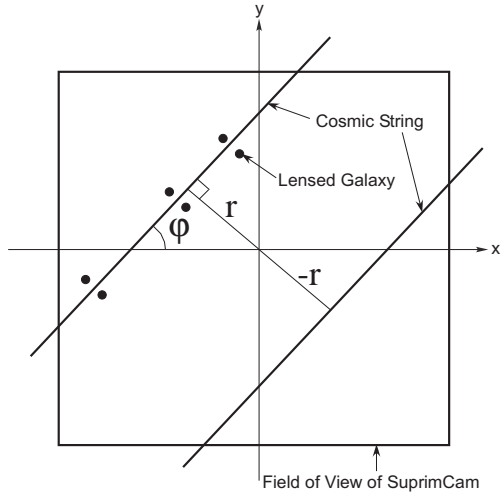}
      }
   }
   \parbox[b]{0.5\textwidth}{
      \caption{Definition of string configuration parameters $r$
               and $\phi$.}
      \label{fig:cs-search}
   }
\end{figure}

%
%
%
%
%
%
Our strategy presented here is dedicated to searching for a straight
string.
In this case, double images lensed by the string are aligned along
it, so the strategy is simply to find such aligned objects
which have a pair of similar brightness, color, and morphology
separated to the same direction within 5 arcseconds.
The procedure to find the aligned pair objects is:
(1)  Make an object catalog for each Suprime-Cam image.
(2)  Select objects which have a pair of similar brightness and
     color within 5 arcseconds.
(3)  Calculate the string configuration parameter $r_{i}$ and
     $\phi_{i}$ for each $i$-th pair. These parameters are defined in
     Fig.~\ref{fig:cs-search}.
(4)  Calculate error density function $P_{i}(r,\phi)$.
(5)  Likelihood for a set of parameters ($r$,$\phi$) is obtained by
     taking a sum of $P_{i}(r,\phi)$. 
(6)  Estimate chance probability for the likelihood to be
     greater than the maximum of $P_{i}(r,\phi)$. 
If the chance probability is smaller than 1\%, it indicates the
existence of a straight string.

\section{Simulation Study}

\begin{figure}
   \centerline{
      \parbox[t]{0.5\textwidth}{
         \includegraphics[width=0.45\textwidth]{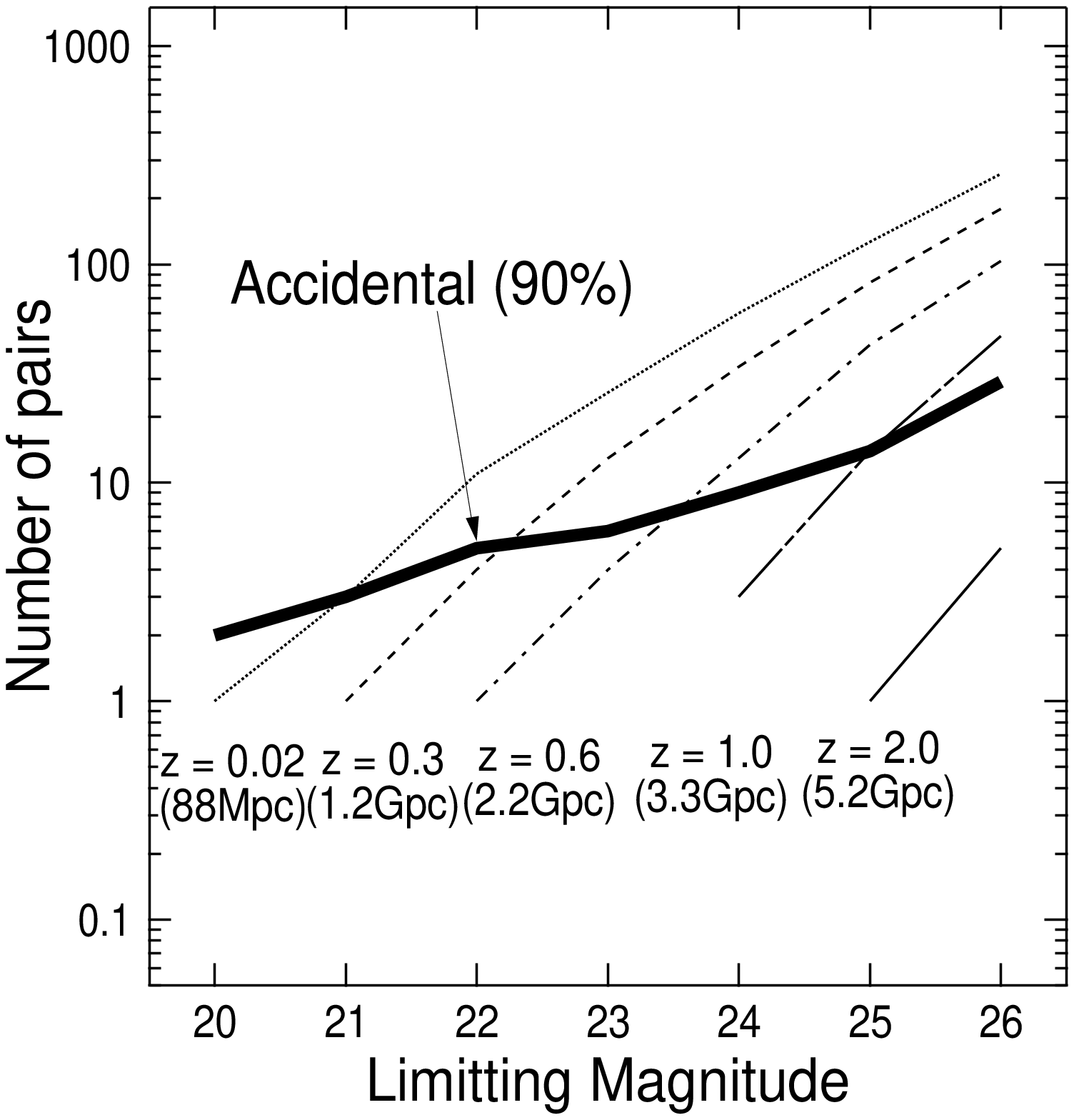}
      }
      \parbox[t]{0.5\textwidth}{
         \includegraphics[width=0.45\textwidth]{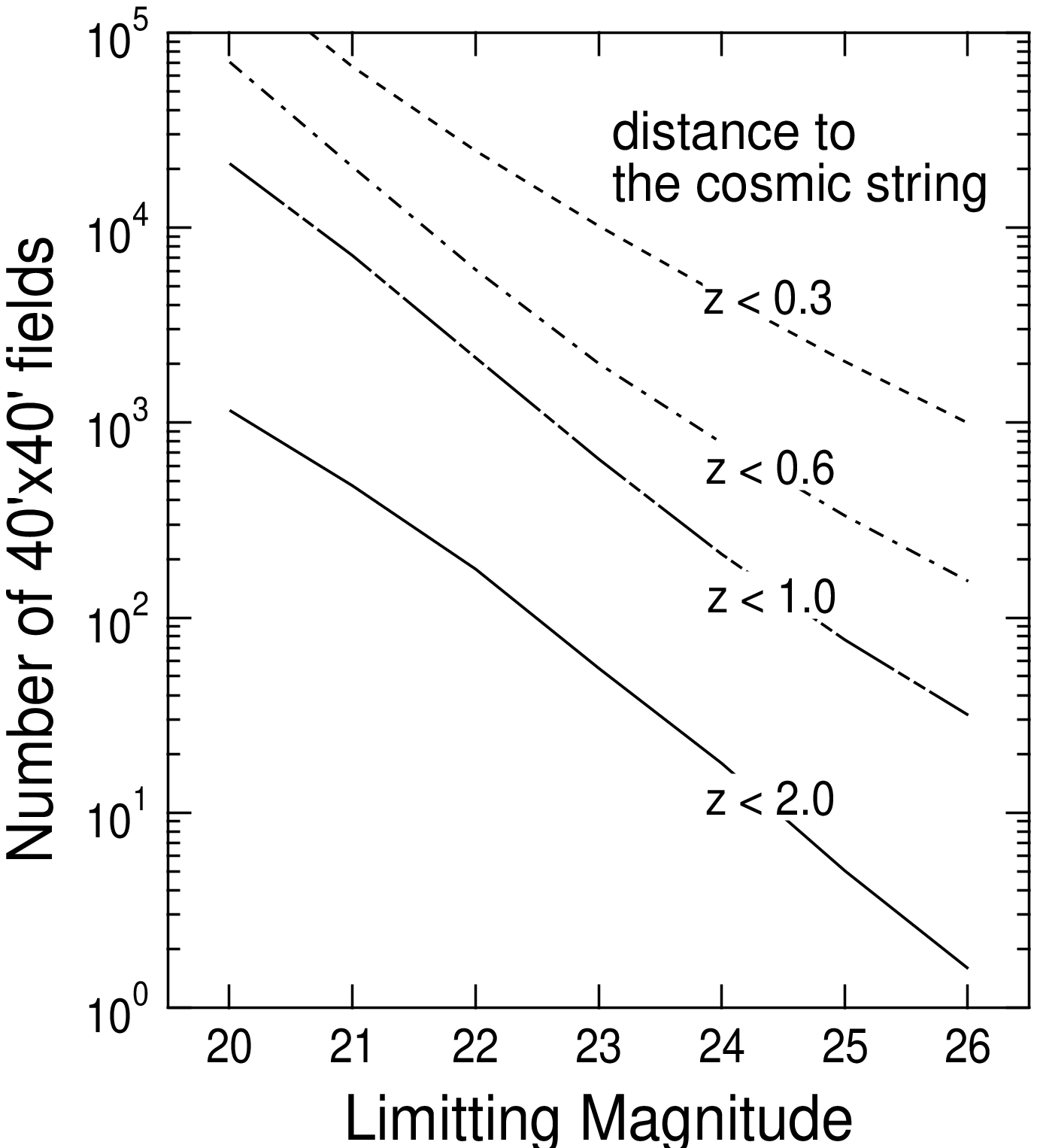}
      }
    }
    \caption{Left panel shows the expected number of lensed images in a
             Suprime-Cam field of view as
             a function of limiting magnitude of the observation. 
             The redshifts of the string are $z = $ 0.02, 0.3, 0.6,
             1.0 and 2.0 from left to right, respectively. 
             The number of accidental alignment of pair galaxies is
             represented by a heavy line.
             Right panel shows the numbers of $40'\times40'$ fields to
             be observed for detecting one string for each
             string redshift.}
     \label{fig:simulation}
\end{figure}

We have estimated probability to detect double images lensed by a
string under the assumption that the total length of the string is 
31 in horizon units~[2] and it is straight at least in the
field of view of the Suprime-Cam.
We used numerical galaxy catalog developed by Nagashima and Yahagi
to estimate the number of galaxy lensed by the string.
In the left panel of Fig.~\ref{fig:simulation}, the numbers of observed
pair galaxies
for each limiting magnitude in R-band are shown for the cases that the
string is at distances $z =$ 0.02, 0.3, 0.6, 1.0 and 2.0.
The numbers of accidental alignment of pair galaxies in 90\% C.L. are
also shown in the figure.
From this figure, one can see that if a string is as close as
$z = 0.02$ ($\sim$ 88 Mpc), observation with 21 limiting magnitude is
enough to detect the cosmic string.
If the string is as far as $z = 1.0$, we need to do an
observation deeper than 25 magnitude.
In the right panel of Fig.~\ref{fig:simulation}, the numbers of
$40^{\circ}\times 40^{\circ}$ field necessary for detecting one string in
average are shown for each limiting magnitude.
For detecting one cosmic string which is at distance $z < 1.0$,
we need 80 Suprime-Cam fields taken under the condition as deep as 25
limiting magnitude.

\section{Application to observation data}

We have applied the string search procedure described above to the
archived Suprime-Cam data.
The data used for this analysis are summarized in Table~\ref{tbl:data}.
In constructing a list of pair objects, the following conditions are
applied to make a pair: the difference of the magnitudes
in R band is less than 1.0 mag and the color difference is less than
0.05 for V$-$R, $i'-$R and $z'-$R, and 0.15 for B$-$R if the data of
each band is available.
The result is shown in Table~\ref{tbl:result}.
The chance probability is calculated by Monte Carlo simulation, that is,
(1) rearrange the observed pairs at a random position in the observed
field in a random direction, (2) calculate the maximum likelihood,
(3) repeat this 1000 times, and then obtain the distribution of maximum
likelihood.
The distribution obtained by the simulation can be regarded as that
for a null-string condition.
We could not find any significant evidence of existence of a long
straight cosmic string in the four fields in 99\%C.L.

\begin{table}
   \begin{center}
   \begin{tabular}{cccc} \hline
      Filed            & Size  & available band and limiting magnitude \\ \hline\hline
      AGASA Field      & 1.8 deg$^{2}$  & R$<$22.5 \\
      SDF              & 0.3 deg$^{2}$  & B$<$26.8, V$<$26.6, R$<$26.7, $i'<26.2, z'<26.2$\\
      SXDF             & 1.3 deg$^{2}$  & B$<$27.4, R$<$27.3, $i'<27.0, z'<26.7$ \\ 
      2 deg$^2$ Field  & 2.2 deg$^{2}$  & R$<$24.4, I$<$24.1\\ \hline
   \end{tabular}
   \end{center}
   \caption{Summary of the data used for this analysis. The AGASA field
           is one of the directions where the EHCR cosmic rays are
           observed by AGASA experiment. 
           }
   \label{tbl:data}
\end{table}

\begin{table}
   \begin{center}
      \begin{tabular}{ccccc} \hline
         Field          & $Likelihood_{max}$ & aligned pairs & total
                          pairs & chance probability \\ \hline \hline
         AGASA Field    &  1.947 &  7 & 3383 &  3.3\%  \\
         SDF            &  0.827 &  3 &   99 & 49.0\%  \\
         SXDS           &  1.642 &  5 & 1214 & 21.2\%  \\
        2 deg$^2$ Filed &  3.322 & 10 & 4578 &  7.1\%  \\ \hline
      \end{tabular}
   \end{center}
   \caption{The result of the straight string search. The column of
            'total pairs' represents the number of pair galaxies which has
            similar brightness and/or color with 6 arcsec separation. 
            The 
            column of 'aligned pairs' represents the number of the
            pairs separated to the same direction and aligned in a
            straight line.
            }
   \label{tbl:result}
\end{table}

\endofpaper
\end{document}